# Segregation-induced changes in grain boundary cohesion and embrittlement in binary alloys

Michael A. Gibson and Christopher A. Schuh[*]

Department of Materials Science and Engineering, Massachusetts Institute of

Technology, 77 Massachusetts Avenue, Cambridge MA 02139 USA

Contact Information:

Michael A Gibson: m_gibson@mit.edu, 617-258-5032

Christopher A Schuh: schuh@mit.edu, 617-253-6901 [*]Corresponding Author

**Grain boundary embrittlement occurs when a solute enriches at a grain boundary and lowers its cohesive energy. While grain boundary enrichment is often attributed to equilibrium segregation effects, most models of embrittlement consider either the energetics of decohesion or the equilibrium adsorption at the boundary, but not both phenomena together. We develop a model for the change in cohesive energy of a grain boundary of a pure metal upon introduction of solute under conditions of equilibrium segregation prior to fracture. A heuristic grain boundary cohesion map is presented to delineate whether a given solute-solvent pair will exhibit weakening or strengthening of grain boundaries. The analysis helps to clarify that grain boundary embrittlement requires a solute that will both lower the cohesive energy of the boundary and segregate to it in the first place. The map reasonably captures known metal-metal embrittling pairs.**

# 1. Introduction

Phenomenologically, metal-metal embrittlement occurs when an otherwise tough metal loses its ability to plastically dissipate large amounts of energy during crack propagation, owing to the presence of a second, metallic embrittling species [1]. Fractographic analyses reveal that such failures often occur along grain boundaries [2, 3], where embrittlement via segregation or wetting of tramp elements has occurred. Such impurity-induced embrittlement has been shown to commonly occur in most metallic base metals of interest, such as Cu- [4], Ni- [5], Fe- [6], and Al-based [7-11] systems. A grain boundary is considered embrittled when boundary cleavage (i.e. breaking the bonds across a grain boundary) becomes energetically favorable to blunting mechanisms (i.e. dislocation emission) [12]. The energetic barrier to cleavage is the grain boundary cohesive energy [13], which is given by:

$$E_{\text{GBC}} = 2\gamma_S - \gamma_{\text{GB}} \qquad (1)$$

where $E_{\text{GBC}}$ is the energy of grain boundary cohesion, $\gamma_S$ is the surface energy and $\gamma_{\text{GB}}$ is the grain boundary energy. The surface and grain boundary energies are both a strong function of the alloy chemistry, since solute atoms segregate to the boundary according to the Gibbs adsorption isotherm [14]:

$$d\gamma = -\sum_i \Gamma_i d\mu_i \qquad (2)$$

where $d\gamma$ is a differential change in interfacial energy, $\Gamma_i$ is the specific excess of solute i at the interface, and $d\mu_i$ is a differential change in the chemical potential of the solute. The strong dependencies of $\gamma_S$ and $\gamma_{\text{GB}}$ on the chemical potential of solutes imply that $E_{\text{GBC}}$ is also a strong function of solution chemistry at equilibrium. As the plastic work needed to extend a crack has been shown to be proportional to $E_{\text{GBC}}^{2-5}$ [15, 16], adsorption effects on material toughness can be unintuitively large.

Hirth, Rice and Wang presented a thermodynamic framework for describing the changes in the cohesive energy of a grain boundary due to solute adsorption by integration of the Gibbs adsorption isotherm under various constraints, formalizing the concept of interfacial cohesive energies under constrained equilibrium [17-20]. Under the constraint of no diffusion of solute to or from the grain boundary during cleavage (i.e.

during fast crack growth relative to diffusion), the following expression was derived for the non-equilibrium work of separation for a given amount of enrichment:

$$E_{GBC}(\Gamma) = E_{GBC}(\Gamma = 0) - \int_0^\Gamma [\mu_{GB}(\Gamma) - \mu_S(\Gamma/2)] d\Gamma \tag{3}$$

where $E_{GBC}(\Gamma = 0)$ is the reversible work of separation of the pure material and $\mu_{GB}$ and $\mu_S$ describe the chemical potentials of the solute at the grain boundary and surface before and after cleavage respectively at the given enrichment, $\Gamma$. The integral in (3) defines the change in grain boundary cohesive energy of the pure material upon addition of solute to the system. In this general treatment, the change in grain boundary cohesion is governed by two quantities: the amount of segregant at the interface, $\Gamma$, and the difference in the chemical potentials at the interface/surfaces before and after segregation, $\mu_{GB}(\Gamma) - \mu_S(\Gamma/2)$. Proper prediction of changes in grain boundary cohesion in specific alloy systems amounts to the simultaneous analysis of these two terms.

Previous modeling work on intergranular embrittlement has mostly explored these two terms independently. The $\mu_{GB}(\Gamma) - \mu_S(\Gamma/2)$ term defines the energetics of embrittlement and was explored first parametrically [21-25], and more recently computationally [22, 26, 27]. Seah used a quasichemical bond-breaking model to describe the embrittling tendency of different solutes in iron [21]. More recently, Geng et al. described embrittling potency in Fe and Ni via a modified bond-breaking model with an added elastic mismatch term, and correlated the predictions with ab-initio results for grain boundary cohesive energies [22, 23]. Additional ab-initio studies have been performed for many other alloy systems; see for example [22, 26, 27]. A recent study by Lejcek and Sob reviewed the existing literature on changes in grain boundary cohesion for Fe-based systems [28]. Their systematic review suggested that changes in grain boundary cohesive energies upon alloying can be quantitatively predicted by the difference between the enthalpies of sublimation of the solute and solvent, in accord with the analytical work of Seah [21].

While these models consider the energetics of the solute at the grain boundary versus the surface (essentially the $(\mu_{GB}(\Gamma) - \mu_S(\Gamma/2))$ term in Hirth, Rice and Wang's treatment), they include limited or no treatment of whether the solute will enrich at the grain boundary to begin with. Grain boundary segregation can be described with surface adsorption analogs [29], as first conceptualized by McLean [13], with one of the most successful theories describing co-segregation and temper embrittlement in multinary iron alloys via the use of multinary interaction parameters [30, 31]. Segregation to grain boundaries can also be studied computationally through Monte Carlo simulations [32-34], but such studies are relatively infrequent and, to the authors' knowledge, have not been explicitly linked to changes in cohesion.

Hence the problem of equilibrium interfacial cohesion is still relatively unexplored. Several groups have treated the problem in the thermodynamic abstract [17-20, 35, 36], while other groups have done so parametrically [36-38]. Recently, Lejcek has applied the Fowler isotherm to a quasichemical, bond-breaking model to simultaneously study the embrittling tendency and equilibrium segregation in both binary [37] and higher order [38] alloy systems. To the authors' knowledge, this is the only study to provide a framework for the quantitative treatment of embrittlement in the presence of equilibrium segregation using readily measureable materials parameters. However, the developments of Lejcek are analytical and parametric, and use only equilibrium properties of interfaces, and are thus not readily applicable to changes in interfacial cohesion due to non-equilibrium segregation [5, 6], or to the chemically-constrained equilibrium of fast fracture. Additionally, the models in [37, 38] are limited to open systems, and are therefore less relevant for, e.g., fine-grained or nanocrystalline alloys, which must be treated as closed systems because segregation may appreciably change the concentration of solute remaining in the bulk.

It is our purpose in this study to develop a model capable of describing both equilibrium and non-equilibrium segregation-induced changes in grain boundary cohesion using readily available materials parameters as inputs. Our developments are in the spirit of Lejcek's works above, but can be applied for fast and slow fracture conditions, include a consideration of the solubility limit and second phase precipitation, and are generalizable to capture both open and closed systems. The model is intentionally built under the same thermodynamic framework as previous models for the thermodynamics of polycrystals to enable a unified description of segregation –induced reductions in both grain boundary energies and grain boundary cohesion. Such a unified description should facilitate the design of nanocrystalline alloys which exhibit reduced grain growth and are not embrittled by segregation. We also use the model to systematically screen almost 2000 binary pairs from the periodic table, to provide guidance for future alloy development.

**2. Quasichemical Bond Breaking Model**

We begin by adapting the definition of $E_{\text{GBC}}$ from Eq. (1) to represent the difference in internal energy of the crystal in the presence of a grain boundary versus the internal energy of the crystal with the same atomic configuration, but with two free surfaces replacing the grain boundary:

$$E_{\text{GBC}} = E_{\text{surfaces}} - E_{\text{grain boundary}} \qquad (4)$$

This quantity represents the work that must be provided by an external agent to create a free surface at a grain boundary, neglecting plastic deformation. The cohesive energy of a pure material, A, is then described by Eq. (1). The change in cohesion upon alloying, $\Delta E_{\text{GBC}}$, is given by the difference between the cohesive energies in the alloyed and the unalloyed states:

$$\Delta E_{\text{GBC}} = E_{\text{GBC}}^{\text{alloy}} - E_{\text{GBC}}^{\text{pure}} \qquad (5)$$

Ignoring irreversible processes, a positive $\Delta E_{\text{GBC}}$ represents a net increase in the thermodynamic resistance to decohesion, while a negative $\Delta E_{\text{GBC}}$ represents a tendency for embrittlement [17].

Inspired by the success of Guttmann and Seah's models [21, 30] in predicting temper embrittlement via use of interaction parameters, we propose a model for $\Delta E_{\text{GBC}}$ of a binary alloy which employs interaction parameters appropriate to the bulk, surface, and intergranular regions of an alloy. Such a formulation is analogous to that used in several previous treatments of nanocrystalline stability [39-44], but has not previously been used to examine free surface effects or embrittlement. We hope that this formulation may be used to provide a unified description of embrittlement and thermal stability. We specifically consider the "regular nanocrystalline solution" model, which is a regular solution model based on nearest neighbor interactions in a polycrystalline structure [20], for which the internal energy is given as:

$$E^{gb}_{solution} = \sum_{r=gb,b,t} N_r^{AA} E_r^{AA} + N_r^{BB} E_r^{BB} + N_r^{AB} E_r^{AB} \quad (6)$$

Here $E_r^{XY}$ represents a pairwise bonding energy between atoms of species X and Y in the region r and $N_r^{XY}$ represents the total number of bonds between atoms of X and Y in region r. The regions, r, are captured by the subscripts gb, b, and t, corresponding to grain boundary, bulk, and transitional regions, respectively. The regions are permitted to have distinct bonding energies, i.e. $E_{gb}^{AA} \neq E_b^{AA}$, and distinct compositions. The energies of the transitional bonds are assumed to be the same as those of the bulk bonds in this work.

The internal energy of a solution containing a surface can be similarly defined:

$$E^{s}_{solution} = \sum_{r=s,b,t} N_r^{AA} E_r^{AA} + N_r^{BB} E_r^{BB} + N_r^{AB} E_r^{AB} \quad (7)$$

where the subscript s now refers to the surface region. A schematic representing the geometry of the grain boundary and surface regions being considered is shown in Fig. 1.

We calculate $\Delta E_{GBC}$ by assuming random mixing in each of the bulk and interfacial regions and introducing the resulting numbers of bonds into Eqs. 6 and 7, to obtain

$$N_r^{XY} = N_r \cdot P(XY|r) \quad (8)$$

Where $N_r$ is the total number of bonds in region $r$ and $P(XY|r)$ is the probability of a randomly chosen bond being an X-Y bond given that we are selecting bonds from region $r$. Under the assumption of random mixing, the $P(XY|r)$ are only functions of the local solute concentration in the various regions: $X_{gb}, X_s$, and $X_b$. The values for the various $N_r$ and $P(XY|r)$ are given in Table 1. The parameter ν is the transitional bond fraction, and represents bonds at the grain boundary that span between the bulk and grain boundary atoms. A value of ν = ½ is appropriate for a monolayer of atoms in an FCC lattice with the interface oriented along the close-packed plane, and is a value used in previous models developed in the same framework [41]. The bond counting is performed on a per-area basis such that all of the following free energies follow the convention of defining $E_{GBC}$ on the basis of energy per unit area of surface or interface. The bond energies are normalized by the quantity $\frac{zt_{gb}}{2\Omega}$ to account for their areal bond densities, with $z$ being the coordination of the solvent, $t_{gb}$ the grain boundary thickness (taken as 0.5 nm in this study [45]), and $\Omega$ the atomic volume of the solvent [41].

If we assume that fracture takes place rapidly relative to diffusion, the bonding in the bulk regions will not change, and $E_{GBC}^{alloy}$ can be calculated only from considering the changes in bond energies in the region near the interface. In Fig.1, this corresponds to only considering the regions with blue bonds; the bonds directly adjacent to the interface. The fast-fracture analysis corresponds to an upper bound on the change in cohesive energy of the grain boundary: any equilibration of the surface must correspond to a lowering of the surface energy, and thus a reduction in cohesive energy. Thus, any solutes predicted to embrittle a host must increase the degree of embrittlement upon equilibration of the surface, while any solutes predicted to increase the cohesion of a GB must exhibit a decrease in grain boundary cohesion upon equilibration of the surface.

Introducing the quantities in Table 1 into Eqs. (7) and (8) with $X_{gb} = X_s$, we arrive at the following expression for the internal energies of the near-grain boundary and near-surface regions:

$$E_{solution}^{gb} = \frac{z\,t_{gb}}{2\Omega}(1-v)\left[X_{gb}E_{gb}^{BB} + (1-X_{gb})E_{gb}^{AA} + 2X_{gb}(1-X_{gb})\omega_{gb}\right] \quad (9)$$

$$E_{solution}^{s} = \frac{z\,t_{gb}}{2\Omega}(1-2v)\left[X_{gb}E_{s}^{BB} + (1-X_{gb})E_{s}^{AA} + 2X_{gb}(1-X_{gb})\omega_{s}\right] \quad (10)$$

where the grain boundary and surface interaction parameters are defined as:

$$\omega_{gb} = E_{gb}^{AB} - \frac{E_{gb}^{AA}+E_{gb}^{BB}}{2} \quad (11a)$$

$$\omega_s = E_s^{AB} - \frac{E_s^{AA}+E_s^{BB}}{2} \quad (11b)$$

$\omega_{gb}$ and $\omega_b$ can be shown to be related to the dilute heat of grain boundary segregation, $\Delta H_0^{seg}$, and the dilute heat of mixing, $\Delta H^{mix}$, by following the algebraic manipulations of [41] to arrive at:

$$\Delta H_0^{seg} = z\left(1-\frac{v}{2}\right)\left[\omega_b - \omega_{gb} - \frac{\Omega}{zt}\left(\gamma_{gb}^B - \gamma_{gb}^A\right)\right] \quad (12a)$$

$$\Delta H^{\text{mix}} = z\omega_b \tag{12b}$$

We note that there is a difference between Eq 12a and its analogue in Ref [41]. This is due to a typographical error that is corrected in the present work, where Ref. [41] included an additional factor of two in the definition of ν (Eq. 10 in Ref. [41]). The related equations in this work can be produced following the mathematical steps as Ref. [41] with the factor of two removed. The energetics of grain boundary segregation, such as the elastic mismatch energy between the solute and solvent, are thus incorporated into the bonding energies of the model as detailed in the appendix.

Introducing Eqs. (9) and (10) into Eq. (4), $E_{\text{GBC}}^{\text{alloy}}$ is then given by the difference between the internal energies of the surface and grain boundary:

$$E_{\text{GBC}}^{\text{alloy}} = \frac{z\,t_{\text{gb}}}{2\Omega}\left((1-\nu)\left[X_{\text{gb}}\Delta E_{\text{gb}\to s}^{BB} + (1-X_{\text{gb}})\Delta E_{\text{gb}\to s}^{AA} + 2X_{\text{gb}}(1-X_{\text{gb}})\Delta\omega_{\text{gb}\to s}\right] - \frac{\nu}{2}\left[X_{\text{gb}}E_s^{BB} + (1-X_{\text{gb}})E_s^{AA} + 2X_{\text{gb}}(1-X_{\text{gb}})\omega_s\right]\right) \tag{13}$$

where we denote the differences in the energetic parameters, $E$, between the surface and grain boundary states by $\Delta E_{\text{gb}\to s}$, for example, the difference between the surface and grain boundary interaction parameters is written as $\Delta\omega_{\text{gb}\to s} = \omega_s - \omega_{\text{gb}}$. It is conventional in quasi-chemical models of surface energy to assume that $E_s^{AA} = E_b^{AA}$ and $E_s^{BB} = E_b^{BB}$, i.e. that like bond energies near a surface are the same as their bulk counterparts [21, 24, 25, 46]. Under this condition, Eq. (13) reduces to:

$$E_{\text{GBC}}^{\text{alloy}} = \frac{z\,t_{\text{gb}}}{2\Omega}\left((1-\nu)\left[X_{\text{gb}}(E_b^{BB} - E_{\text{gb}}^{BB}) + (1-X_{\text{gb}})(E_b^{AA} - E_{\text{gb}}^{AA}) + 2X_{\text{gb}}(1-X_{\text{gb}})\Delta\omega_{\text{gb}\to s}\right] - \frac{\nu}{2}\left[X_{\text{gb}}E_b^{BB} + (1-X_{\text{gb}})E_b^{AA} + 2X_{\text{gb}}(1-X_{\text{gb}})\omega_s\right]\right) \tag{14}$$

We then make the following substitutions:

$$\gamma_{\text{gb}}^i = \frac{z\,t_{\text{gb}}(1-\nu)}{2\Omega_i}\left(E_{\text{gb}}^{ii} - E_b^{ii}\right) \tag{15}$$

$$2\gamma_s^i = -\frac{\nu z\,t_{\text{gb}}}{4\Omega_i}E_b^{ii} \tag{16}$$

and subtract the cohesive energy of the solvent, A, to obtain the change in cohesive energy due to the presence of a solute at the grain boundary:

$$\Delta E_{\text{GBC}}^{\text{alloy}} = X_{\text{gb}}[(2\gamma_s^B - \gamma_{\text{gb}}^B) - (2\gamma_s^A - \gamma_{\text{gb}}^A)] + \frac{z\, t_{\text{gb}}}{\Omega} X_{\text{gb}}(1 - X_{\text{gb}})[(1 - \tfrac{3\nu}{2})\Delta\omega_{\text{gb}\to s} - \tfrac{\nu}{2}\omega_{\text{gb}}] \quad (17)$$

This expression properly obeys the limit of the cohesive energy of the GB being that of either pure A or B at $X_{\text{gb}} = 0$ or $1$, respectively. Assuming $\Delta\omega_{\text{gb}\to s} = 0$ (i.e. that the bond energies between unlike atoms do not change significantly between the surface and grain boundary states), $\nu = 1/2$, and $\gamma_{\text{gb}}^B = \gamma_s^B/3$ [47] (See also the appendix) yields the following simple expression for the cohesive energy of the grain boundary, which will be used in the rest of this paper:

$$\Delta E_{\text{GBC}}^{\text{alloy}} = 5X_{\text{gb}}/3[\gamma_s^B - \gamma_s^A] - \frac{z\, t_{\text{gb}}}{4\Omega} X_{\text{gb}}(1 - X_{\text{gb}})\omega_{\text{gb}} \quad (18)$$

This expression for $\Delta E_{\text{GBC}}$ contains two types of terms. The first is a bond-breaking term, corresponding to a rule-of-mixtures expression for the change in cohesive energy of the grain boundary upon addition of solute based on the elemental grain boundary cohesive energies. The second is a bond-breaking mixing term proportional to the grain boundary interaction parameter, indicating whether A-B interactions at the grain boundary are favored at the grain boundary, and thus whether such interactions strengthen or weaken the boundary; comparison with Eq. (17) reveals a second-order correction to this term that accounts for the difference in grain boundary and surface interaction parameters. This second-order correction can be used to account for additional relaxations at the surface.

Eq. (18) is the baseline model output of this work, describing local changes in grain boundary cohesion upon alloying, but without any thermodynamic consideration of whether or not the same alloy would in fact adopt a grain boundary-segregated state in equilibrium. In this sense, the model is similar to that proposed by Seah [21], except that the present model presupposes the presence of a grain boundary, and is defined in the context of a model for describing equilibrium segregation [39-44]. Later in this paper we will take advantage of this feature of the model to explore the effects of equilibrium segregation explicitly.

**3. Alloy Pair Screening Without Thermodynamic Segregation**

In order to study the changes in grain boundary cohesive energy for specific alloy systems, we use a database of thermodynamic parameters to calculate the change in the cohesive energy of a grain boundary upon alloying for 1978 binary alloy pairs. The values used for the surface energies and mixing enthalpies are described in the appendix, and the full set of numerical data used is collected in the online supplement.

We begin by examining the relative importance of the different energetic contributions to $\Delta E_{\text{GBC}}$ by artificially setting the composition at the grain boundary, and studying the resulting changes in cohesion for specific alloy pairs. Because the cohesive energy of a grain boundary upon alloying should be considered relative to its cohesive energy in the pure state, all the plots below represent this information using the ratio $E_{\text{GBC}}^{\text{alloy}}/E_{\text{GBC}}^{\text{pure}} = \frac{\Delta E_{\text{GBC}}}{E_{\text{GBC}}^{\text{pure}}} + 1$. For purposes of illustration, it is useful to consider three quantities that capture physically unique aspects of the alloy's bonding energetics at the grain boundary. First is the ratio of the surface energies of the solute compared to the solvent, $\gamma_s^B/\gamma_s^A$, which expresses the relative strength of the solute atoms' bonds compared to those of the solvent atoms. Second and third are $\Delta H^{\text{mix}}$ and $\Delta E^{\text{elastic}}$. The latter term is the elastic energy associated with the solute substitution, and is one of the contributors to the mixing energy, as described in the Appendix. These two parameters, $\Delta H^{\text{mix}}$ and $\Delta E^{\text{elastic}}$ determine the value of $\omega_{\text{gb}}$ (via Eq. (12) and the developments in the Appendix), and thus control whether A-B bonds are favorable compared to A-A and B-B bonds alone. The values of $\Delta H^{\text{mix}}$ and $\Delta E^{\text{elastic}}$ thus combine to explain deviations from a rule-of-mixtures type of behavior for the grain boundary cohesive energy. We examine these variables in turn in what follows.

Fig. 2 displays a log-log plot of $E_{\text{GBC}}^{\text{alloy}}/E_{\text{GBC}}^{\text{pure}}$ versus $\gamma_s^B/\gamma_s^A$ evaluated at grain boundary compositions of $X_{\text{gb}} = 0.3$ and $0.75$ for each solvent-solute pair. A solute-solvent pair is considered embrittling if $E_{\text{GBC}}^{\text{alloy}}/E_{\text{GBC}}^{\text{pure}}$ is less than one, i.e. if the point representing the alloy pair lies below the x-axis in Fig. 2. Panel **2A** with $X_{\text{gb}} = 0.3$ represents the case of a moderately enriched boundary. A clear positive correlation exists between the normalized cohesive energy of the alloyed grain boundary and the ratio of the surface energies. This correlation is strengthened for a strongly enriched boundary, shown in 2**B**, as should be the case for a boundary composed primarily of B atoms. In both 2**A** and 2**B**, alloying with a more cohesive alloying element, i.e. pairs with $\gamma_s^B/\gamma_s^A > 1$, rarely results in a decrease of the grain boundary cohesive energy.

Deviations from rule-of-mixtures behavior occur due to chemical interaction between the solute and solvent at the GB. As the interaction parameter is a function of $\Delta H^{\mathrm{mix}}$ and $\Delta E^{\mathrm{elastic}}$, these deviations are determined by the values of these quantities for each alloy pair. In Fig. 2, the points on the plots are colored according to the value of the Miedema heat of mixing [48]. The color gradient across the width of the distribution demonstrates that the majority of the scatter in the data can be attributed to the value of the heat of mixing. The dark blue tails on the distributions show that couples with the most positive heats of mixing tend to be those with the largest differences in surface energies.

In order to better illustrate how the heat of mixing and elastic mismatch energy affect predictions of the change in cohesion upon alloying, $E_{\mathrm{GBC}}^{\mathrm{alloy}}/E_{\mathrm{GBC}}^{\mathrm{pure}}$ is plotted against these two variables in Fig. 3 for a grain boundary composition of $X_{\mathrm{gb}} = 0.3$. The couples in both of these panels are colored according to the value of $\gamma_s^B/\gamma_s^A$ for each couple. Panel 3**A** shows that negative heat of mixing couples are unlikely to suffer a decrease in grain boundary cohesive energy for moderately enriched boundaries. This is intuitive, as the A-B bonds gained upon alloying and grain boundary segregation should be stronger than A-A bonds alone, and from the above discussion, we can expect the B-B bonds to be approximately as strong as the A-A bonds in negative heat of mixing couples. For positive heat of mixing alloys, the $E_{\mathrm{GBC}}^{\mathrm{alloy}}/E_{\mathrm{GBC}}^{\mathrm{pure}}$ distribution forks based upon the position of $\gamma_s^B/\gamma_s^A$ relative to a value of unity, as indicated by the coloring. Again, this can be intuitively related to bond energies; if the alloying element is more cohesive than the base element, i.e. $\gamma_s^B/\gamma_s^A > 1$, the blue-colored, upward fork in Fig. 3**A** indicates that the change in GB cohesion will generally be positive, despite the fact that A-B bonds are weaker than A-A or B-B bonds on average. However, if both A-B and B-B bonds are weaker than A-A bonds, $\gamma_s^B/\gamma_s^A < 1$ and $\Delta H^{mix} > 0$, then the grain boundary will be strongly embrittled, as shown by the red-colored, downward fork in 3**A**.

We note that the plot of $E_{GBC}^{alloy}/E_{GBC}^{pure}$ vs. $\Delta E^{elastic}$ in 3**B** does not display any strong trends. Although a large $\Delta E^{elastic}$ implies a more favorable interaction parameter via Eq. 12a, this is not reflected in the data for two reasons. First, $\Delta E^{elastic}$ is generally much smaller in magnitude than the mixing energies and elemental cohesive energies. Second, $\Delta E^{elastic}$ positively co-varies with the heat of mixing; couples with a large elastic mismatch should display a more positive heat of mixing.

## 4. Comparison with Previous Models

It is instructive to compare the present model to previous models for grain boundary embrittlement to understand its similarities and differentiating characteristics. As was previously mentioned, the form of the change in cohesive energy previously derived is nearly the same as that of Seah [21]. We now compare the present quasichemical model to that of Geng, Freeman, and Olson [22], a phenomenological model derived to explain changes in cohesive energy upon solute addition to grain boundaries as calculated from density functional theory (DFT). Their model contains a bond breaking term and an elastic mismatch term, and is therefore similar to the present model in that both take into account the effects of elastic mismatch energy, the bulk heat of mixing, and the relative bonding strength of the solute relative to the solvent.

Fig. 4 shows 'periodic table graphs' of the data from Geng et al. and similar plots developed using the present model with $X_{gb} = 0.9$. The agreement between Fig 4a and 4b shows conformity between the present model and that of Geng et al. Because this good matching is attained at highly enriched boundaries (where elastic and interaction-based effects can be neglected), the main contribution to the predicted embrittling potencies of Geng et al. is the difference in the cohesive energies of the solute and solvent, which is in agreement with the conclusions of Seah [21]. We further correlate the predictions of our analytical model with ab-initio results on changes in grain boundary cohesive energies upon introduction of a substitutional solute [22, 49]. Figure 5 demonstrates the close match between such simulations and our analytical results, again calculated for highly enriched grain boundaries with $X_{gb} = 0.9$.

**5. Effect of Equilibrium Segregation on Alloy Screening**

Quantifying the embrittling potency of an alloying element if it has segregated to the boundary, as done in the section above, is a necessary first step in explaining embrittlement trends. However, a more relevant analysis would consider the simultaneous prediction of whether impurity elements will segregate to the grain boundary in the first place. For purposes of illustration, the consideration of equilibrium segregation is divided into two steps: first, an isotherm is applied to consider changes in segregation for a given composition. Next, the systems are constrained to be at or below their solubility limits as dictated by the regular solution model. We demonstrate that both constraints are necessary to understand trends in equilibrium segregation.

As we have assumed regular solution interactions, the logical choice for a grain boundary segregation isotherm is a Fowler-like isotherm. As shown by Guttmann and McLean [50], equating the chemical potential at a regularly interacting grain boundary with a regularly interacting bulk system yields an isotherm of the form (written in the notation of the present study):

$$\frac{X_{gb}}{1-X_{gb}} = \frac{X_b}{1-X_b} \exp\left[-\frac{\Delta G_o^{seg} + z\omega_{gb} - z\omega_b - 2(X_{gb}z\omega_{gb} - X_b z\omega_b)}{RT}\right] \quad (19)$$

where $\Delta G_o^{seg}$ is some composition-independent energy released upon segregation (proportional to the difference in grain boundaries energies in our model). Assuming that the interaction parameters at the grain boundary and bulk are equal simplifies the isotherm to the same form as the classical Fowler isotherm [51]. However, Eq. 19 is distinct in form and spirit from the Fowler isotherm, as the original treatment was derived in the context of a substance adsorbing on a surface from a gas phase. Hence, the equilibrium modeled by Fowler is between an ideally interacting solution, a gas, and a regularly interacting one at a surface, as opposed to two regularly interacting solutions as in the case of grain boundary segregation. Consequently, even though one can achieve an isotherm of the same form as the original Fowler isotherm by assuming that the interaction parameters are equal, the interaction parameter $z\omega_b$, often denoted by α in other studies, is not an independent fitting parameter, but is dictated by the mixing behavior of the bulk solution. Our choice of notation reflects this connection with bulk mixing thermodynamics.

The changes in the normalized grain boundary cohesive energies upon application of Eq. 19 are shown in Fig. 6 for all transition metal alloy pairs with a bulk composition $X_b = 10^{-4}$ typical of tramp elements and a temperature of T = 300 K. Fig. 6**A** is analogous to Fig. 3**A**, while Fig. 6**B** is analogous to Fig. 2**B**.

Earlier in reference to Fig. 3A and Fig 2, we saw that the heat of mixing of an alloy system had a small contribution to embrittlement relative to surface energy differences between the solute and solvent. By comparing the previous figures with Fig. 6A, we now see that upon application of chemical equilibrium via an isotherm, $\Delta H^{mix}$ has a very large effect on cohesive energies; the heat of mixing of the alloy pair influences its tendency to segregate substantially, which in turn indirectly affects its tendency to exhibit embrittlement. In particular, we see that if $\Delta H^{mix} < 0$, i.e. if the couple interacts favorably in the bulk, the cohesive energy of the grain boundary is unaffected, as there is no change in the composition of the grain boundary. This lack of segregation due to favorable mixing thermodynamics is the reason that many less-cohesive elements do not strongly embrittle more cohesive hosts, with one example being Al in Fe.

The coloring of the points in 6**A** demonstrates that if the couple does exhibit a positive heat of mixing and thus segregates, then the degree of embrittlement is largely determined by the ratio $\gamma_B^s/\gamma_A^s$; if an impurity is present at a grain boundary, the boundary's change in cohesion is dictated by the strength of the impurity's bonds relative to its host. This trend is made more explicit in Fig. 6**B**, where the ratio of surface energies is represented explicitly on the x-axis. There are two essentially separate trends in Fig. 6**B**: a horizontal line of points along the value $E_{GBC}^{alloy}/E_{GBC}^{pure} = 1$, and a second diagonal line of points with a slope of unity along the line $E_{\text{GBC}}^{\text{alloy}}/E_{\text{GBC}}^{\text{pure}} = \gamma_s^B/\gamma_s^A$. The horizontal line corresponds to the case of an alloy where the solute does not enrich at the grain boundary, so the grain boundary's cohesive properties don't change. The diagonal line corresponds to the case of a strongly segregating system with highly enriched boundaries, in which case the grain boundary becomes composed almost entirely of solute atoms instead of solvent atoms. The scatter between these two trends is for systems with a moderate segregation tendency.

The second constraint for chemical equilibrium is that the system only be considered at compositions below its solubility limit; above the limit additional solute is tied up in second phases and not available to affect the grain boundary cohesion via segregation. We model the solubility limit in the regular solution approximation by assuming the temperature is half the melting temperature of the solvent such that solid-state diffusion and equilibrium GB segregation is kinetically possible. We solve for the minimum of the free energy curve as a function of composition for each alloy system, yielding the solubility limit, $X^{\max}$, within the regular solution approximation. If the solubility limit is greater than $X^{\max} = 10^{-3}$, then the system is modeled as possessing a concentration of $X^{\max} = 10^{-3}$. Otherwise, the system is modeled as having a concentration equal to 80% of its solubility limit; this assignment is arbitrary but does not materially affect the outcome of the analysis. Using this new composition, we again solve Eq. 19 for the equilibrium grain boundary concentration, and calculate the change in cohesive energy of the grain boundary that results.

Figs 6**C** and 6**D** show the distributions of changes in cohesive energies upon application of this constraint, plotted on the same axes as Figs 6**A** and 6**B** to ease comparison of trends. As can be seen, the strengthening branch towards the upper right quadrant in 6**A** collapses back onto the x-axis in Fig 6**C**. The observation that the strengthening branch collapses upon application of equilibrium solubility while the embrittling branch remains strongly segregating is a reflection of the competition between the composition-independent heat of segregation (given by the difference in grain boundary energies) and the composition-dependent mixing effects. While these two

effects both favor segregation for couples in the embrittling branch, these two effects oppose one another for the strengthening branch in 6**A**. For the majority of systems, the forces opposing segregation of more cohesive elements to grain boundaries are stronger than those driving segregation to grain boundaries. Hence, this model predicts that while elements that weaken cohesion at grain boundaries should be commonly observed, it should be relatively rare to observe elements that enhance cohesion upon segregation, which is in line with general experimental observations across the literature.

The main lesson from Fig. 6 is that embrittlement only occurs if two conditions are simultaneously met:

1. The alloying element segregates. This condition depends upon temperature, composition, and the energetics of segregation, with low temperatures, enriched compositions, and positive heats of mixing favoring segregation. We have further shown that one must not only consider the energetics of segregation to grain boundaries, but must also consider the effect of solubility limits when modeling the behavior at chemical equilibrium.

2. The alloying element has significantly weaker bonding than the base metal, quantified by $\gamma_s^B/\gamma_s^A < 1$. This condition is influenced primarily by the surface energies of the two pure elements, and is quantitatively supported by the trends observed in Fig. 5.

If a system is at equilibrium with respect to solute segregation, then both of these two conditions are necessary for embrittlement to occur. Whereas previous models have demonstrated that differences in bonding energy such as $\gamma_s^B/\gamma_s^A$ are a measure of the embrittling potency of an alloying element, the above analysis demonstrates that $\Delta H^{mix}$ is equally important, as it is a thermodynamic measure of the solute's tendency to enrich at the grain boundary and is also related to the solubility limit.

**6. Grain Boundary Embrittlement Map**

The above discussion shows that the values of $\gamma_s^B/\gamma_s^A$ and $\Delta H^{mix}$ for a given alloy pair together dictate the degree of grain boundary embrittlement and enrichment at equilibrium. It is instructive to plot these two parameters against one another in order to generate a design map for metal embrittlement. This yields a 'grain boundary cohesion map,' shown in Fig. 7 for all solvent-solute pairs for which data was available; the full set of numerical data used in this chart is provided in the online supplemental material. Here the heat of mixing is non-dimensionalized by the energy scale of the solvent at half the melting point of the base metal, $RT_M^A/2$, because the effects of equilibrium segregation are most apparent when specimens are equilibrated at sufficiently high temperatures that diffusion permits chemical equilibration but low enough that entropic effects do not drive the solute back into solution. The dots representing each alloy pair are colored according to the change in cohesive energy of the grain boundary induced upon alloying calculated from Eq. 18, with $X_{gb}$ calculated under the same conditions as Fig. 6**C**: the solutes are required to be below the solubility limit and the isotherm is enforced at half of the solvent's melting temperature.

The blue, purple, and magenta dots in Fig. 7 indicate progressively larger proportional reductions of grain boundary cohesion upon alloying. The distribution of these dots shows that only couples in the top left quadrant exhibit large proportional reductions in grain boundary cohesion; a positive heat of mixing and a lower cohesive energy are required for embrittlement to occur. We term this the 'embrittling quadrant.' However, not all couples within this quadrant are brittle under the conditions specified. In particular, couples with sufficiently small heats of mixing and large ratios of $\gamma_s^B/\gamma_s^A$ in the embrittling quadrant are not expected to cause significant embrittlement, because they do not have a strong enough segregation tendency in the first place; these points lie on an envelope near the x- and y-axes, and are colored green in Fig. 7. As previously noted, the details of the segregation behavior of each solute-solvent pair are a function of the constraints applied to the system. For example, applying the isotherm (Eq 19) at a lower homologous temperature will generally predict larger grain boundary enrichment, and thus larger proportional changes in grain boundary cohesion. We find that the alloy pairs which lie to the top left in this map are generally alloy pairs that one would expect to embrittle; they are mostly refractory BCC metals (i.e. W, Os, Ta, Nb) alloyed with simple metals and semimetals (K, Na, Sr, Tl, Pb, Bi). The reader is referred to the online supplement for all of the data in tabular format.

Under the assumptions of the present model, the remaining three quadrants of Fig. 7 do not exhibit significant grain boundary enrichment at equilibrium. However, this does not mean that the alloy pairs in these quadrants are not of practical interest. As demonstrated in Fig. 2, the right half of Fig. 7 contains alloys that, if solute were present at the grain boundary, would tend to strengthen grain boundary cohesion rather than weaken it. An example of an anti-segregating pair is Fe alloyed with Mo ($\frac{\gamma_s^{Mo}}{\gamma_s^{Fe}} = 1.3, \Delta H^{mix} = 19$ kJ/mol, $E_{GBC}^{Fe+Mo}/E_{GBC}^{Fe}$=1.001), which is used in steels to prevent the segregation of P to grain boundaries due to favorable interactions in the bulk [52, 53]. This prediction arises naturally from our model, but we note that it is different from previous analyses of GB cohesion, which generally predict Mo to strengthen Fe grain boundaries [49]. However, at high enough P enrichment, Mo can be attracted to GB's via favorable interactions with P. In such ternary segregation scenarios, Mo presence is associated with a de-embrittling effect at nearly constant P enrichment [54]. This can be interpreted as a grain boundary strengthening effect on the part of the Mo [49], and shows that one must explicitly account for the expected degree of enrichment in modeling changes in grain boundary cohesion.

The lower left quadrant exhibits similar phenomenology; it features alloys where, even if solute is present at the boundary, it tends not to induce large reductions in grain boundary cohesion (see the left half of Fig. 3**A**). As such, the alloy pairs outside of the embrittling quadrant can be considered 'safe' pairs; the solvent is not predicted to lose cohesion upon alloying, even if non-equilibrium segregation is induced.

We note that there is an analogy between Figs. 7 and the cohesion map of Seah (Fig. 4 of ref. [21]), who presented a similar alloy screening tool for iron alloys plotting the elemental heat of sublimation (on the y-axis) and the lattice parameter (on the x-axis). Both of these axes were developed by Seah to capture solely the energetics of grain boundary decohesion (i.e., neglecting equilibrium grain boundary segregation tendency). Based on our developments in Eqs. 17-19, we can now replace the heuristic axes of Seah with the more general presentation of Figs. 7 and 8. Despite the differences in approach between our map and that of Seah, the physical meanings of our axes and his are closely related. For example, Seah's heat of sublimation is correlated to our surface energy (the y-axes on the respective maps), as both quantities are reflective of the intrinsic strength of a bulk bond. While Seah did not discuss equilibrium segregation, the lattice parameter, his x-axis, may be related to equilibrium segregation in two ways. First, segregation is driven in part by relief of elastic mismatch energies and second, in a related effect, the difference in atomic radii is known to have a large effect on solubility, and thus grain boundary enrichment. Our use of $\Delta H^{\text{mix}}/RT_{\text{M}}$ is in line with the empirical relationship between grain boundary enrichment and bulk mixing thermodynamics [55, 56].

One can re-plot the results of Fig. 7 for smaller subsets of points in order to facilitate a closer examination of its predictions, as done in Fig. 8 for Fe-based alloys. It is fairly well-established that embrittling agents in Fe-based systems include Zn, Ge, Sn, P, As, Sb, Bi, S, Se, Te, and Mn [1]. We see that most metals that are known to embrittle iron are contained in the appropriate 'embrittling quadrant.' Also in line with the predictions of this theory are that the metallic pairs predicted not to embrittle Fe generally do not do so. For example, common alloying elements in steels such as Cr, Ni, Al, Mo and V are known not to embrittle Fe grain boundaries. The agreement between the map predictions and the collective knowledge of iron embrittlement is thus very good. A few discrepancies are present, but these are all easily rationalized by the inaccuracy of Miedema-based heats of mixing and regular solution theory (e.g., Mn and the semimetals P, As, Si, and Ge are all poorly captured by the present approach). Somewhat unexpectedly, the most deleterious elements to GB cohesion (and the cohesion of most other transition metals) are all predicted to be simple metals: Cs, Rb, K, etc. However, these elements are so sparingly soluble in Fe that they do not exhibit a significant effect on the mechanical properties of Fe, and we are not aware of quantitative studies of their effects.

A second case study can be conducted on Au-based alloys, for which a comprehensive and consistent set of tensile tests was done by Roberts-Austen in one of the first studies of grain boundary embrittlement [57], which found that while almost all semimetals and K caused grain boundary embrittlement of Au, no transition metal elements studied caused significant reductions in ductility. Our model correctly predicts that Au will not be embrittled by any transition metal elements, but mixing data was not available for semimetals in Au (see the supplementary material on Au-based alloys for details) to investigate the effects of the semimetals on Au cohesion.

In order to visually compare the distribution of embrittling versus 'safe' pairs, 25 known metal-metal segregating, embrittling pairs were identified and are listed in Supplementary Table S1. For 11 of these pairs, data was available to identify approximate positions of these pairs on the map. In order to compare the distribution of these embrittling elements to elements that are known not to cause reductions in grain boundary cohesion, major alloying elements in structural Ti-, Zr-, Fe-, and Ni-based alloys were identified from common structural alloys and are listed in Supplementary Table S3. The embrittling and safe sets of binary alloys are plotted as red and blue dots in Fig 9. As can be seen, these two sets of alloy pairs occupy distinct quadrants of the map, with the embrittling pairs falling in the top left quadrant, and the safe alloy pairs occupying the other three.

## 7. Conclusion

A quantitative model for the change in cohesion of a grain boundary of a pure metal upon introduction of a solute has been developed, and unlike most prior models of grain boundary embrittlement, is combined on the same thermodynamic basis with an analysis of the equilibrium composition of solute at the grain boundaries. The result is a prediction of metal-metal embrittlement trends under the conditions of equilibrium grain boundary segregation. The model has been applied to all metal-metal alloy pairs for which data could be obtained.

A comparison of the present model to previous of models of embrittlement potency showed that almost all of the energetic predictions for embrittlement potency are explained by the difference in cohesive energies of the solute and solvent or related, correlated quantities. Accordingly, the importance of the ratio of bonding energies between the solute and solvent, captured by the ratio of their surface energies, $\gamma_s^B/\gamma_s^A$, was emphasized in the development of the present model.

The present model is differentiated from past models in that it considers the expected equilibrium degree of enrichment at the grain boundary. Consistent with the general theory of Rice, Hirth, and Wang, a substantial amount of solute at the grain boundary is a prerequisite for embrittling behavior. Further, we have shown that the additional constraint that the bulk solution itself be at equilibrium with respect to second phase precipitation has a substantial effect on the predictions of equilibrium segregation, with positive heat of mixing, anti-segregating alloys never predicted to enrich at the grain boundary. This emphasizes the need for careful application of the concept of chemical equilibrium in describing grain boundary embrittlement.

An alloy design map was developed by plotting the ratio of the bond energies, $\gamma_s^B/\gamma_s^A$, against a non-dimensionalized heat of mixing, allowing simultaneous consideration of equilibrium segregation and embrittling tendency in alloy design. The expected equilibrium changes in grain boundary cohesion were evaluated for all transition metal alloy pairs for which data was available. Examples from the metallurgical literature show that the phenomenological predictions of the design map are relevant to practical use. The distribution of segregating, embrittling species this design map was shown to be distinct from those of common alloying elements in structural metals, and to obey the predicted trends from the model.

**Acknowledgements:**
We would like to thank Dr. Kisub Cho for assistance in collecting and assessing CALPHAD data. MAG gratefully acknowledges support from the Department of Defense (DoD) through the National Defense Science & Engineering Graduate Fellowship (NDSEG) Program, with primary support from the U.S. Office of Army Research under grant W911NF-14-1-0539, and additional support from the Solid State Solar–Thermal Energy Conversion Center (S3TEC), an Energy Frontier Research Center funded by the US Department of Energy, Office of Science, Office of Basic Energy Sciences under Award Number DE-SC0001299.

**Appendix:**

We assume that the dilute, composition-independent heat of segregation is due to both elastic and chemical effects, i.e. that $\Delta H_0^{\text{seg}} = \Delta E_{\text{elastic}} + \frac{\Omega}{zt}\left(\gamma_{\text{gb}}^{\text{B}} - \gamma_{\text{gb}}^{\text{A}}\right)\left(1 - \frac{v}{2}\right)$. This means that the Friedel elastic mismatch energies were used to provide the difference between the bulk and grain boundary interaction parameters in accordance with Eq. 12a: $z\omega_{\text{b}} = z\omega_{\text{gb}} - \Delta E_{\text{elastic}}/\left(1 - \frac{v}{2}\right)$. The Friedel elastic mismatch energies are the same as those in Ref. [58]. A sensitivity analysis was performed to see how incorporation of elastic effects changes the conclusions of the model. We find that incorporation of elastic effects as a composition-independent free energy of segregation (as in Refs [43, 44]) has only a marginal effect on the systematic trends observed in this study. This is because the heat of mixing is the primary driver of segregation under the assumptions of this study, and positive heat of mixing systems tend also to have large elastic mismatch energies, i.e., to be insoluble.

**Table 1**: Summary of bond counting in the near-surface region

| Region | Number of bonds per unit area | Number of Bonds | Bond Energy | Probability |
|---|---|---|---|---|
| Grain boundary (GB) | $N_{gb} = \dfrac{zt_{gb}}{2\Omega}(1-\nu)$ | $N_{gb}^{AA}$ | $E_{gb}^{AA}$ | $(1-X_{gb})^2$ |
|  |  | $N_{gb}^{BB}$ | $E_{gb}^{BB}$ | $X_{gb}^2$ |
|  |  | $N_{gb}^{AB}$ | $E_{gb}^{AB}$ | $2X_{gb}(1-X_{gb})$ |
| Surface (s) | $N_s = \dfrac{zt_{gb}}{2\Omega}(1-\dfrac{3}{2}\nu)$ | $N_s^{AA}$ | $E_s^{AA}$ | $(1-X_{gb})^2$ |
|  |  | $N_s^{BB}$ | $E_s^{BB}$ | $X_{gb}^2$ |
|  |  | $N_s^{AB}$ | $E_s^{AB}$ | $2X_{gb}(1-X_{gb})$ |
| Transitional (t) | $N_t = \dfrac{zt_{gb}}{2\Omega}\nu$ | $N_t^{AA}$ | $E_{gb}^{AA}$ or $E_s^{AA}$ | $(1-X_b)(1-X_{gb})$ |
|  |  | $N_t^{BB}$ | $E_{gb}^{BB}$ or $E_s^{BB}$ | $X_b X_{gb}$ |
|  |  | $N_t^{AB}$ | $E_{gb}^{AB}$ or $E_s^{AB}$ | $X_b(1-X_{gb}) + X_{gb}(1-X_b)$ |

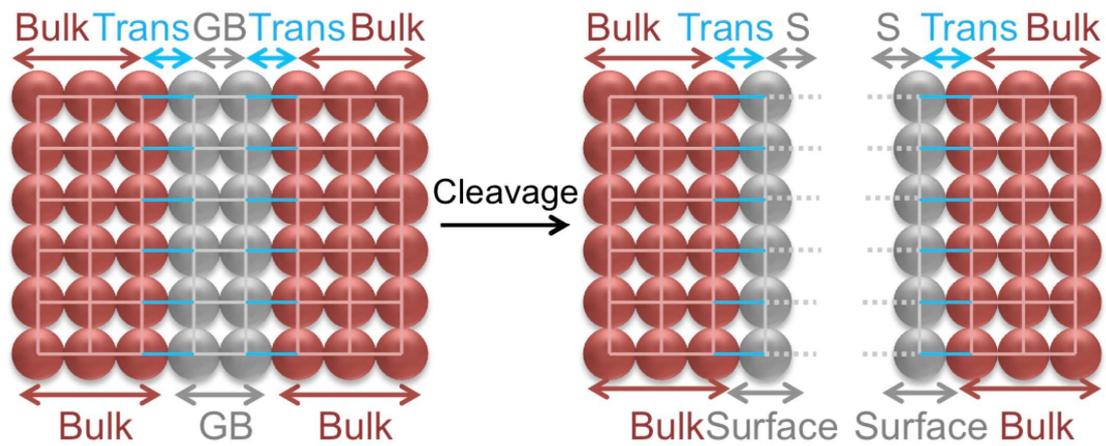

**Figure 1:** Left: The geometry of the Regular Nanocrystalline Solution (RNS) model, which is used to model grain boundary segregation in a regular solution. Right: Geometry of a corresponding bond-breaking model during fracture of the grain boundary.

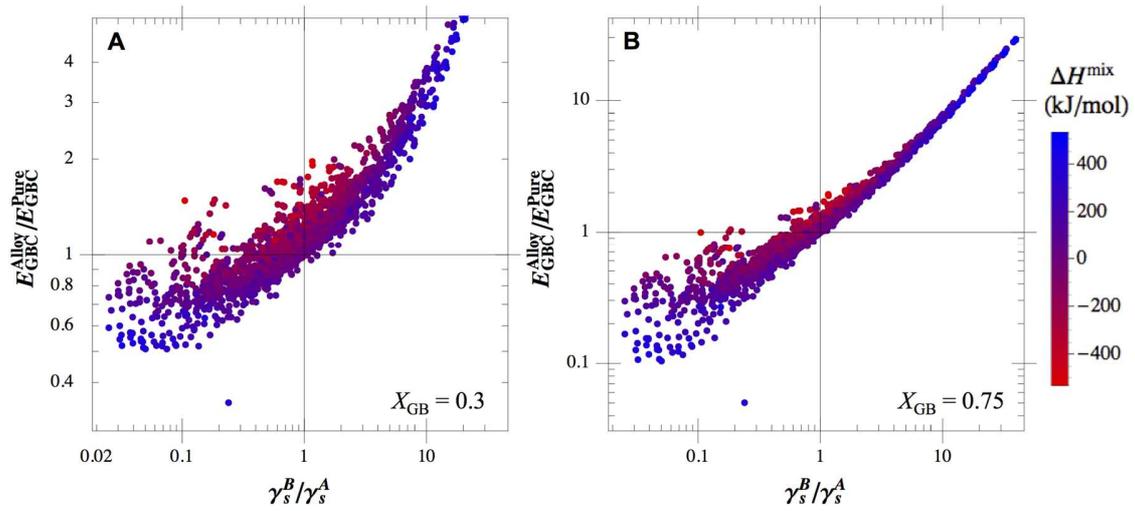

**Figure 2:** A double logarithmic plot of the ratio of the grain boundary cohesive energy upon alloying, $E_{GBC}^{alloy}$, divided by the grain boundary cohesive energy in the pure state, $E_{GBC}^{A}$, versus the ratio of surface energies of the solute and solvent, $\gamma_s^B/\gamma_s^A$, for all transition metal alloy couples plotted for (**A**) $X_{gb} = 0.3$ and (**B**) $X_{gb} = 0.75$. Each point is colored according to the dilute Miedema enthalpy of mixing between the two elements. The full set of numerical data from this plot is available in the online supplement.

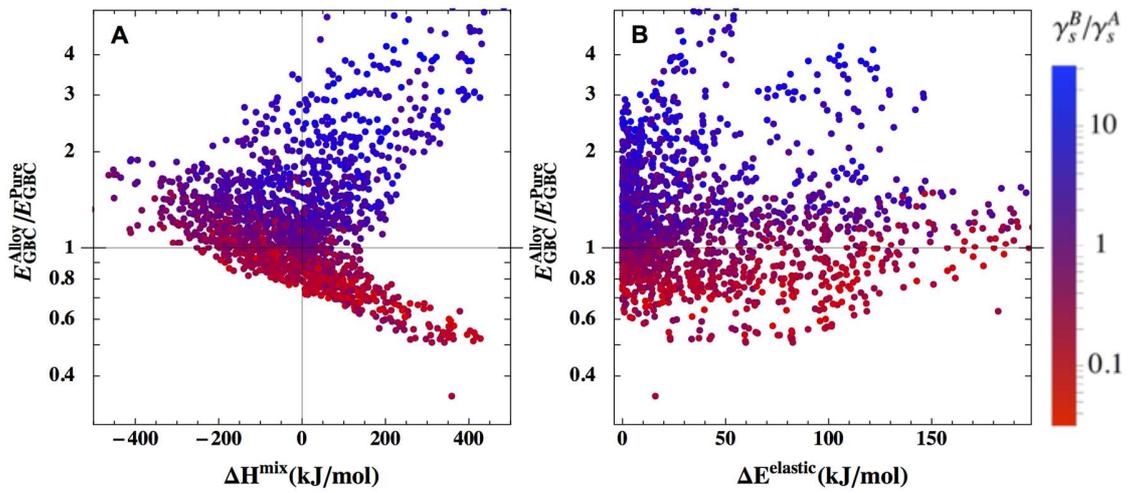

**Figure 3:** The dependence of the change in grain boundary cohesion for all solvent-solute pairs on (**A**) the heat of mixing and (**B**), the elastic mismatch energy for a grain boundary concentration of $X_{gb} = 0.3$. The points are colored according to the log of the ratio of the surface energies of the solvent and solute. The full set of numerical data from this plot is available in the online supplement.

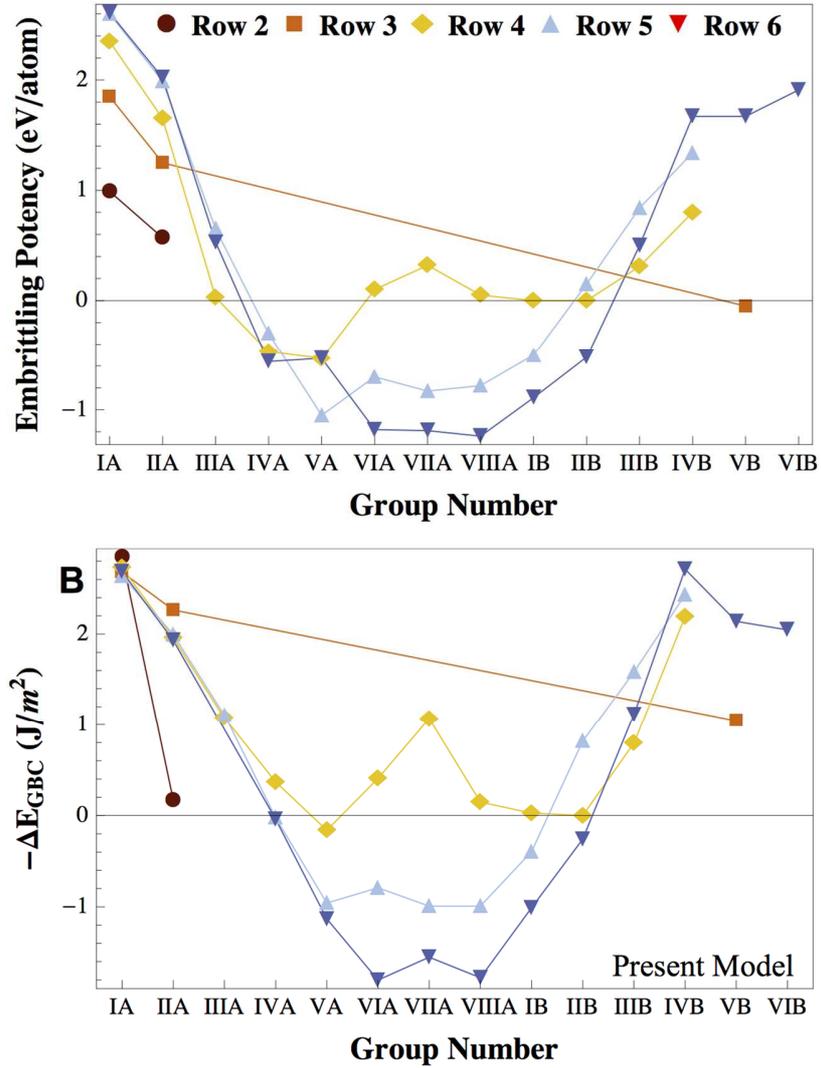

**Figure 4:** (**A**) Reproduction of the 'periodic table graph' for Ni embrittlement of [22], which plots the predicted embrittling potencies as a function of row and group number in the periodic table. (**A**) can be compared with (**B**) which plots the corresponding $\Delta E_{\text{GBC}}$ values from the model of the present study for a system with a grain boundary concentration of $X_{\text{gb}} = 0.9$ as a function of group and row number. Each colored line corresponds to a different row of the periodic table, labeled at the top of (**A**).

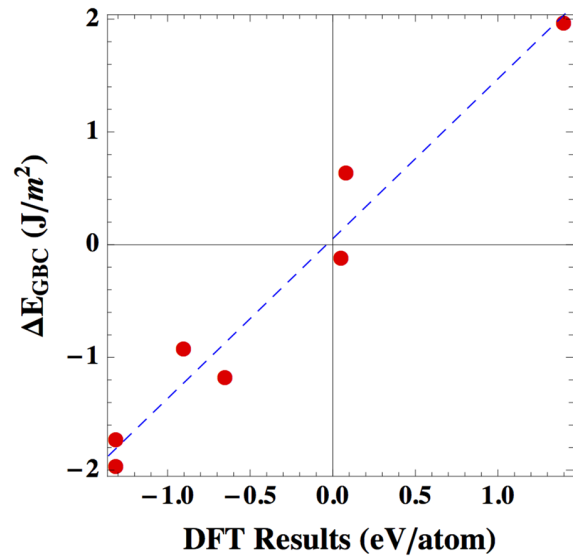

**Figure 5:** Correlation of the present model with density functional theory (DFT) results for a grain boundary concentration of $X_{GB} = 0.9$ Line of best fit shown in blue (adjusted R²=0.95). DFT results are from the Fe-Co, Fe-Ru, Fe-W, Fe-Re, Fe-Mo, Fe-Pd, and Ni-Ca systems [22, 49].

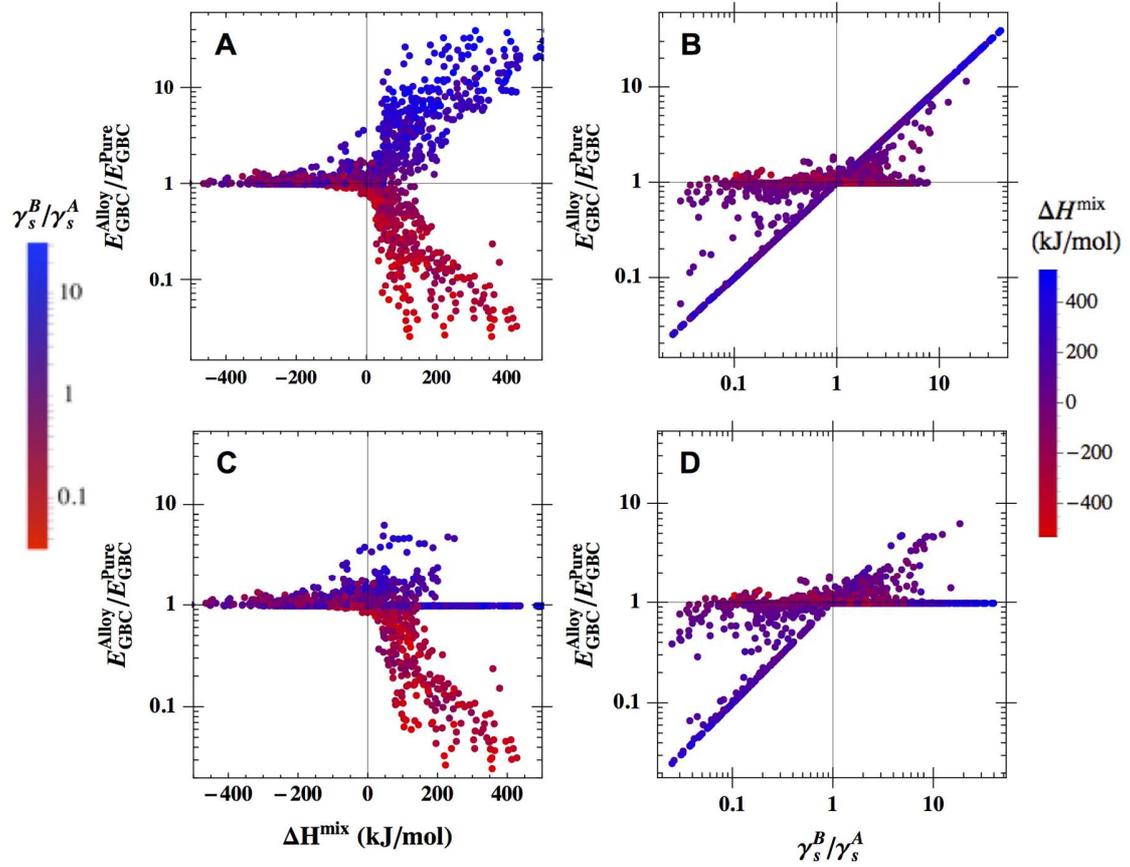

**Figure 6:** The effects of applying different constraints of equilibrium segregation to grain boundary embrittlement, using the same data as Figs. 2 and 3, under the assumption of regular mixing. The ratio of grain boundary cohesive energies with and without solute at the grain boundaries are plotted against Miedema's dilute heat of mixing (**A** and **C**) and the ratio of surface energies (**B** and **D**). In **A** and **B** The system is assumed to be in chemical equilibrium with the bulk at a fixed concentration of $X_b = 10^{-3}$ and T = 300 K. In **C** and **D** the solute is present up to its solubility limit in the bulk. Application of each constraint reduces the number of systems with significant changes in grain boundary cohesion upon alloying. Points are colored according to the ratio of their surface energies on the left, and their heats of mixing on the right.

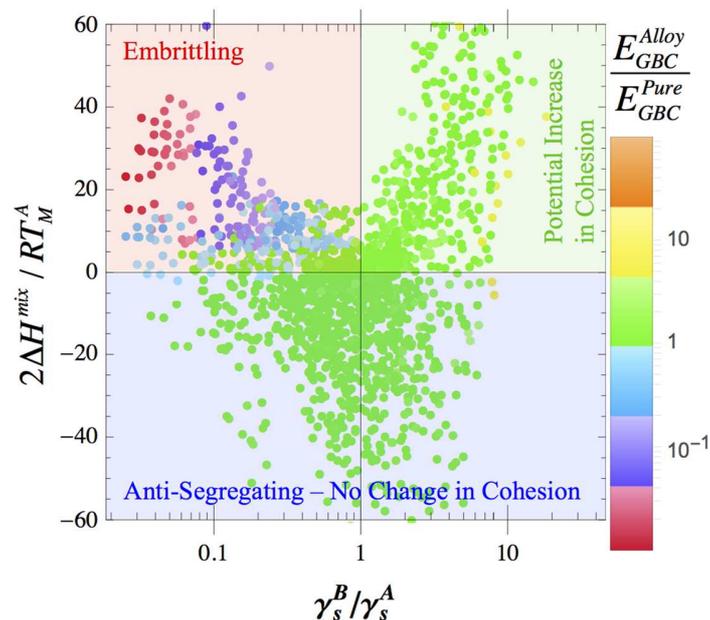

**Figure 7:** A grain boundary cohesion map. Alloy pairs are placed upon the map according to the heat of mixing normalized by the melting temperature of the solvent and the ratio of the surface energy of the solute to the solvent. The pairs are colored according to the predicted ratio of the grain boundary cohesive energy in the alloyed state as compared to the unalloyed state. Cohesive energies are calculated in the same manner as for Figs 6**C** and **D**. The shading and labeling of the different quadrants indicate expected behavior upon alloying. The full set of numerical data from this plot is available in the online supplement.

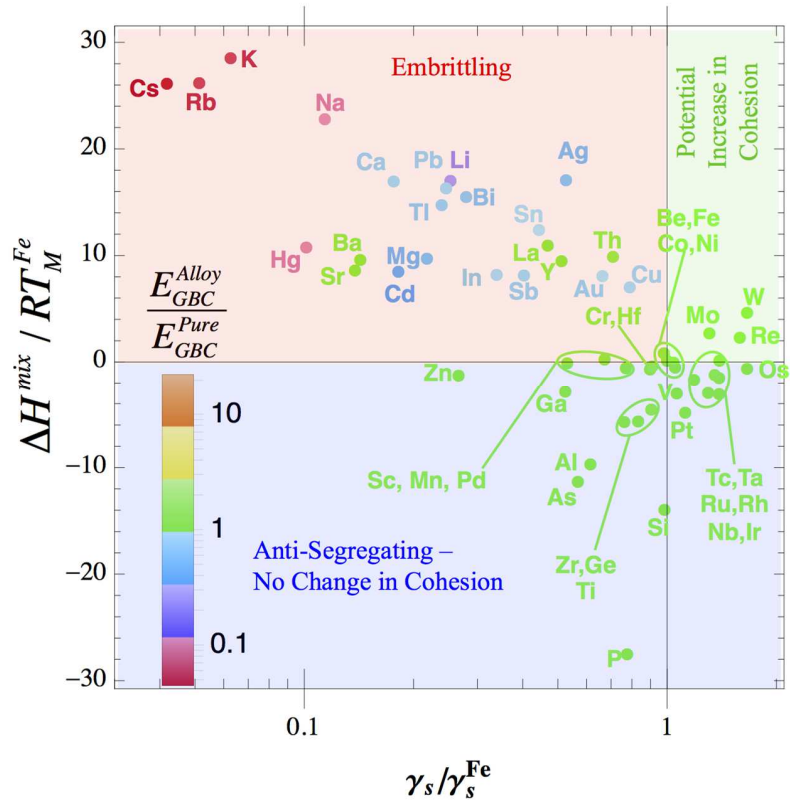

**Figure 8:** The same map as in Figure 7, but only for Fe-based alloys to facilitate comparison with experiment.

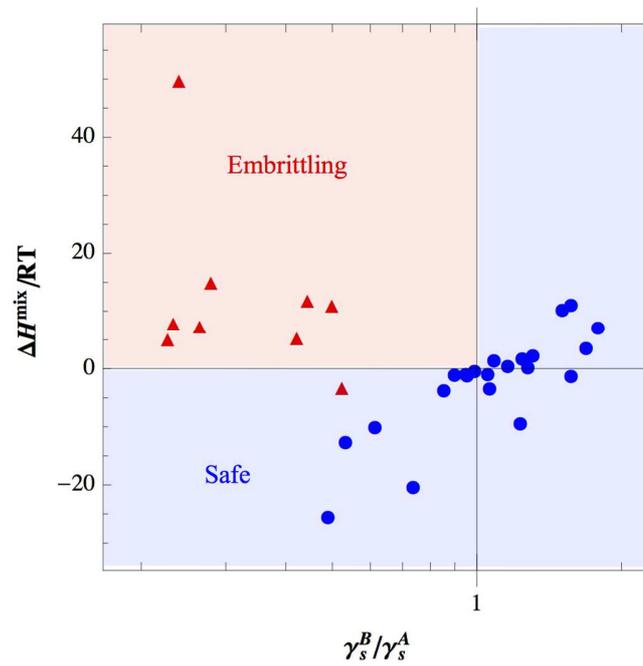

**Figure 9:** A comparison of the distribution of various alloying element behaviors on the same axes as the grain boundary cohesion maps of Figs 7 and 8. The red dots are alloying pairs that exhibit segregation-induced grain boundary failure. The blue elements are safe alloying elements as judged by common alloying elements in structural metals.

# Supplementary Materials For
**Grain boundary cohesion and embrittlement in alloys**
Michael A Gibson and Christopher A Schuh*
*Corresponding author. E-mail schuh@mit.edu

**This file includes:**
1. Experimental evidence for embrittling versus non-embrittling pairs
2. Regression analyses and adjustments of energetic inputs
3. Quantitative validation of Miedema's model for binary mixing enthalpies
4. Cohesion map for gold alloys
5. Supplementary references

**1. Experimental Evidence for Embrittling versus non-embrittling pairs**

We conducted a literature survey to find alloy pairs that have been proven to exhibit segregation-induced changes in toughness. BCC metals such as Fe and W yield straightforward interpretation of embrittlement tendencies, as they tend to exhibit low-temperature GB fracture, and the influence of alloying elements is reflected in changes in ductile to brittle transition temperatures. FCC metals do not exhibit ductile to brittle transitions like BCC metals, and rarely exhibit grain boundary fractures except under extreme embrittlement, such as Cu alloyed with Bi. However, FCC metals do exhibit GB-mediated fracture at high temperatures via creep cavitation, which is a strong function of impurity content as well. If the impurities have been found to segregate to GBs prior to creep cavity formation and have been found to severely reduce creep rupture life or ductility, they have also been included. It is often difficult to distinguish segregation-based embrittlement at high temperatures from embrittlement due to liquid metal embrittlement (LME) or solid-metal (induced) embrittlement (SM(I)E). Pairs for which we were not able to cleanly rule out LME or SME have been omitted from inclusion in Figure 9. Table S1 provides the list of known metal-metal and metal-semimetal segregation-based embrittling pairs used to produce Figure 9.

**Supplementary Table S1:** Pairs that segregate to and embrittle grain boundaries.

| Host | Embrittler | Brief Notes | References |
|------|------------|-------------|------------|
| Ag | Pb | | [1] |
| Ag | Sn | | [1] |
| Ag | Bi | | [2] |
| Al | Li | | [3-5] |
| Al | Zn | Zn can also enhance LME. | [6-9] |
| Au | K | | [10] |
| Au | Bi | | [10] |
| Au | Pb | | [10] |
| Au | Sn | | [10] |
| Au | Tl | | [10] |
| Ni | Bi | | [11-13] |
| Ni | Pb | | [12-14] |

| Ni | Sn | | [13] |
|---|---|---|---|
| Ni | Tl | | [13] |
| Ni | Li | | [15] |
| Ni | Ag | | [12] |
| Cu | Bi | | [16] |
| Cu | Pb | | [17] |
| Cu | Sn | | [18] |
| Fe | Bi | | [19] |
| Fe | Ga | Difficult to find mechanical tests on Galfenol, but limited reports show ~1% ductility and intergranular fracture. | [20, 21] |
| Fe | Sn | | [22] |

Many other alloys may exhibit segregation-based embrittlement, but these were not included in the final analysis due to conflicting claims regarding the mechanism of embrittlement. Some of these pairs are shown in Table S2 to demonstrate the search criteria.

**Supplementary Table S2:** Pairs that do not unambiguously segregate to and embrittle grain-boundaries

| Host | Embrittler | Discussion | References |
|---|---|---|---|
| Al | Na | Other Alkali metals in Al were also not included due to the discussion in the following Refs. | [23-25] |
| Al | Pb | | [26-28] |
| Fe | Pb | | [29] |
| Fe | Cu | | [30] |
| Fe | Zn | | [28] |
| Fe | Mn | Mn has been reported to embrittle in some cases, but is intentionally added to prevent P embrittlement. Ferromanganese generally has high enough levels of As (up to ~0.3%) that it is difficult to disentangle the effects of the Mn from As | [31] |

A list of non-embrittling pairs was needed for comparison in Figure 9. Those in Table S3 were chosen because they are common alloying elements in structural metals, or were experimentally observed to not affect ductility. This is not meant to be an exhaustive list of 'safe' pairs, but merely an illustrative list. Au data from [10].

**Supplementary Table S3:** Elements known not to embrittle grain boundaries of other transition metal elements

| Solvent | Solute |
|---|---|

| Au | Ni |
| Au | Pt |
| Au | Co |
| Au | Fe |
| Au | Zn |
| Au | Cd |
| Au | Cu |
| Au | Pd |
| Au | Rh |
| Au | Ag |
| Au | Al |
| Au | Li |
| Fe | Cr |
| Fe | Ni |
| Fe | Al |
| Fe | V |
| Fe | Mo |
| Ni | Fe |
| Ni | Cr |
| Ni | Co |
| Ni | Mo |
| Ni | Nb |
| Zr | Sn |
| Ti | Al |
| Ti | V |
| Ti | Sn |
| Ti | Zr |
| Ti | Mo |
| W | Re |

**2. Regression analyses and adjustments of energetic inputs**

*2.1 Correlation of grain boundary energy with surface energies for pure metals*

It is commonly stated that the grain boundary energy for a metal is approximately one third of the surface energy, with some evidence of such a correlation tabulated by Murr [32]. In order to quantitatively verify this relationship, we gathered data from [32-34] and correlated the grain boundary energies with the surface energies, requiring a one-to-one correspondence, $\gamma_{GB} = m\,\gamma_S$. As can be seen below, the correlation between the two is good, especially given the experimental difficulties in measuring the surface and grain boundary energies and their sensitivity to alloy purity and environment. The slope of this regression is equal to one third within statistical error (m = 0.358±0.026, 95% confidence, adjusted $R^2$ = 0.986). We therefore use this correlation in the paper to set the pure grain boundary energy equal to one third of the surface energy, and then develop a model for the surface energy from better-tabulated values of the cohesive energy to substitute values for when these quantities are not known from experiment.

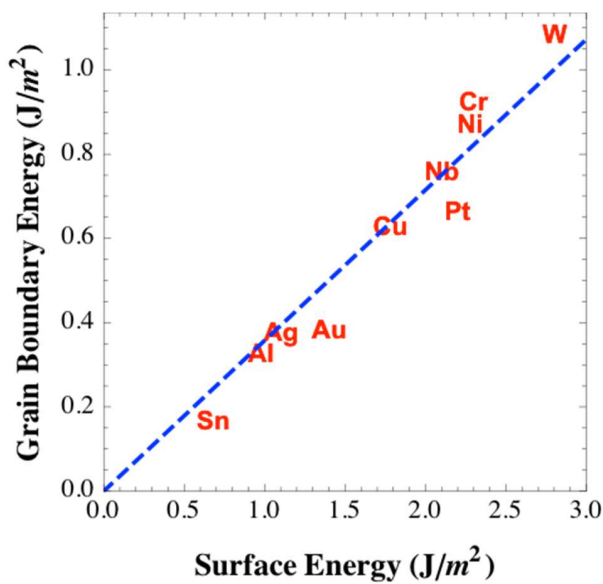

**Figure S1:** Experimentally measured grain boundary energies versus experimentally measured surface energies for a variety of metals showing that the measured grain boundary energy can be approximated as one third of the surface energy.

*2.2 Modeling of surface energies with cohesive energies and validation of model*

Solid-state surface energies and grain boundary energies are not available for all elements of interest. In order to generate a consistent data set, the surface energy was modeled via a bond-breaking model. We assume 1/3 of the bonds are broken at a free surface, such that the energetic penalty of surface creation corresponds to one third of the elemental cohesive energy per atom [35]. This energetic penalty per atom was then normalized by the elemental molar area [35, 36] to give a term with units of $J/m^2$. We found that these estimates overestimated the surface energies by ~ 22%, consistent with electronic and structural relaxation effects at the surface. We thus empirically adjusted the estimates by 22%, with the results of the regression analysis shown below (adjusted $R^2$ for the single parameter regression is 0.981):

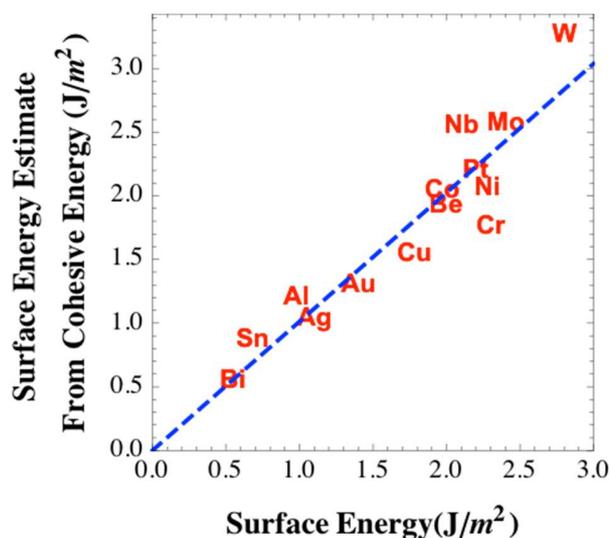

**Figure S2:** Estimated surface energies using cohesive energies and molar areas versus experimentally determined surface energies for a variety of metals.

**3. Quantitative validation of Miedema's model for binary mixing enthalpies**

      A set of binary interaction parameters was needed to estimate mixing effects in the solid state. Miedema's binary heats of mixing [37, 38] are often used for this purpose [39, 40]. In order to verify that the mixing enthalpies of Miedema are quantitatively useful estimates of mixing enthalpies, empirical CALPHAD mixing data from a set of critically assessed, relatively well-studied binary systems within the SGTE Solution database 4 (SSOL4) from Thermocalc [41] was correlated with Miedema's enthalpies of mixing. More specifically, the derivative of the CALPHAD excess enthalpy of mixing with respect to composition in the dilute limit was compared to the Miedema model's dilute heats of mixing. The dilute heat of mixing was chosen rather than the equiatomic heats of mixing, as the main effect of the mixing enthalpy in this study was not on the energetics of cohesion itself, but rather on determining the amount of solute that segregates, which typically pertain to dilute bulk solutions. We find that Miedema's dilute enthalpies of mixing are fair, albeit imperfect predictors of dilute limits of empirical mixing data, both in the liquid and solid state, as shown in the figures below:

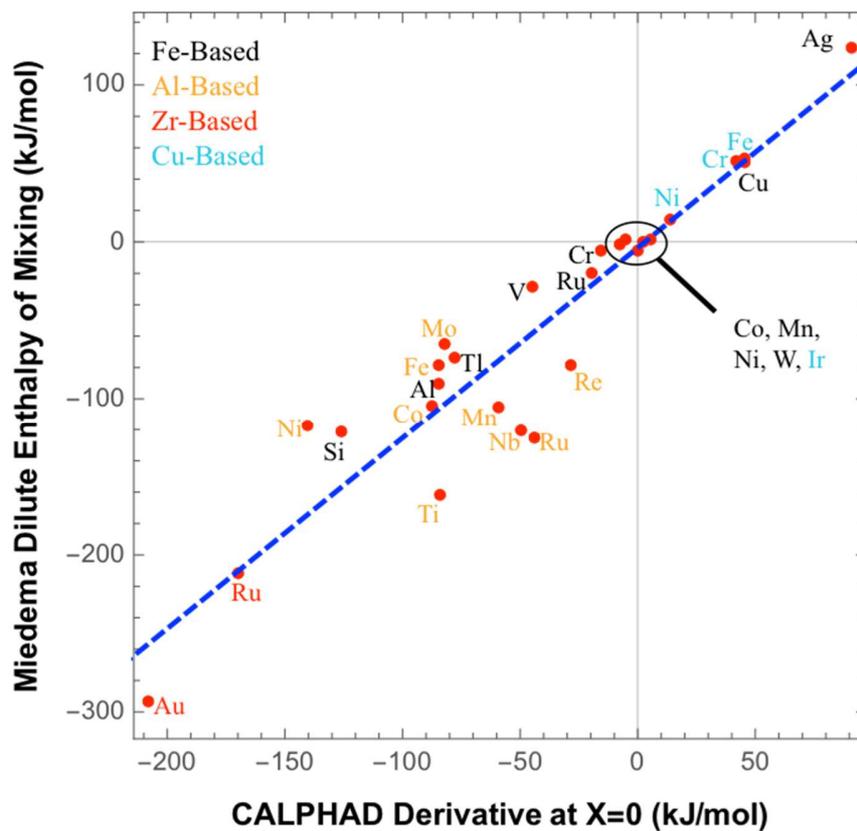

**Figure S3:** Liquid CALPHAD data vs. liquid-based Miedema predicitions. The labels on the points correspond to the solute for each binary pair, and these labels are colored according to the solvent identity. Adjusted $R^2$ for the one parameter regression is 0.913, with the resulting slope being 1.241, so the Miedema model overestimates by ~25% for dilute liquid excess enthalpies of mixing.

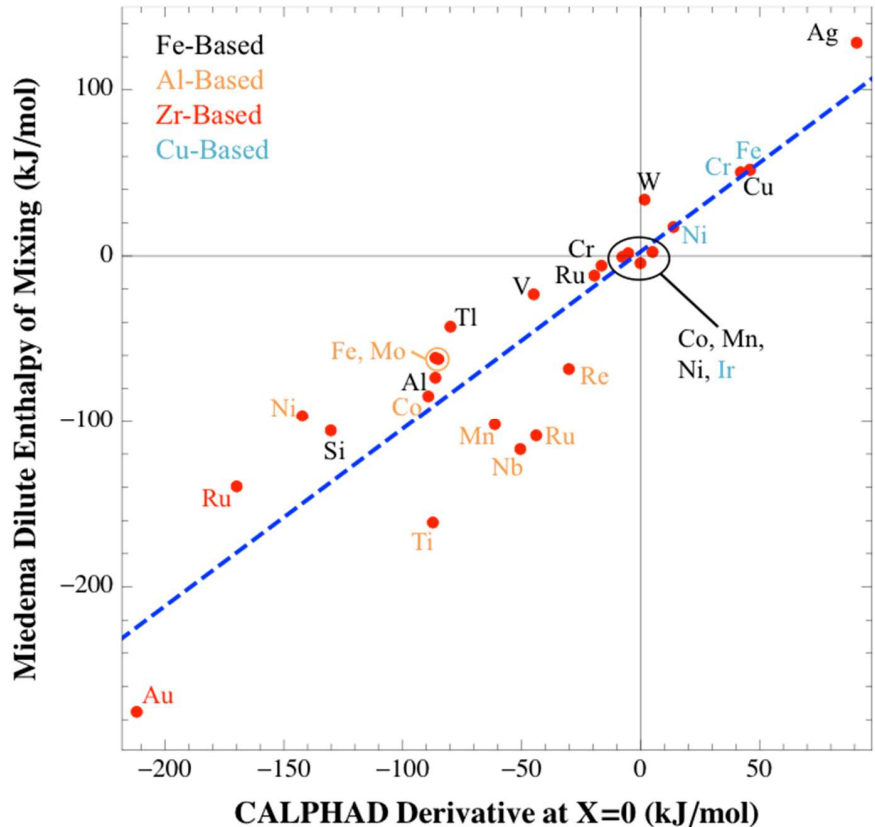

**Figure S4:** Solid CALPHAD data vs. Miedema's crystalline dilute enthalpies of mixing. The labels on the points correspond to the solute for each binary pair, and these labels are colored according to the solvent identity. Adjusted $R^2$ for the one parameter regression is 0.864, with the resulting slope being slope = 1.05±0.16 (95% confidence), so the Miedema model is fairly close to empirical data on average for dilute solid excess enthalpies of mixing.

Figures S3 and S4 demonstrate that the sign of the Miedema heat of mixing is correct for all binaries studied, with the exception of only Fe-Mn, which has a negligibly small error in its heat of mixing and is known to be highly non-regular. This is the important feature of this model, as the sign of the heat of mixing is the major factor in determining enrichment at the grain boundary. As the mixing terms are essentially second-order compared to the surface energies in the modeling the change in cohesion, we conclude that the Miedema model's fidelity in terms of sign and fairly good quantitative prediction of dilute mixing enthalpies are satisfactory for the purposes of this study. In order to be consistent with the corrections made for the surface energy estimates, the heats of mixing were also corrected for electronic relaxation effects at surfaces when calculating changes in cohesive energies by dividing by the same factor of 1.22. The study was done with and without this correction, and the inclusion of this correction did not affect any of the qualitative conclusions of the study, and had only a minor effect on the quantitative values.

## 4. Cohesion Map for Gold-based Alloys

A grain boundary cohesion map (the analog of Figures 7 and 8 in the main text) was also created for Au-based alloys to facilitate comparison with the data from Roberts-Austen. Enthalpies of mixing were not available for semimetals and metalloids in Au, but was available for transition metals alloyed with Au. As can be seen, none of the metals identified as non-embrittling in [10] and listed in Table S3 are predicted to be embrittling.

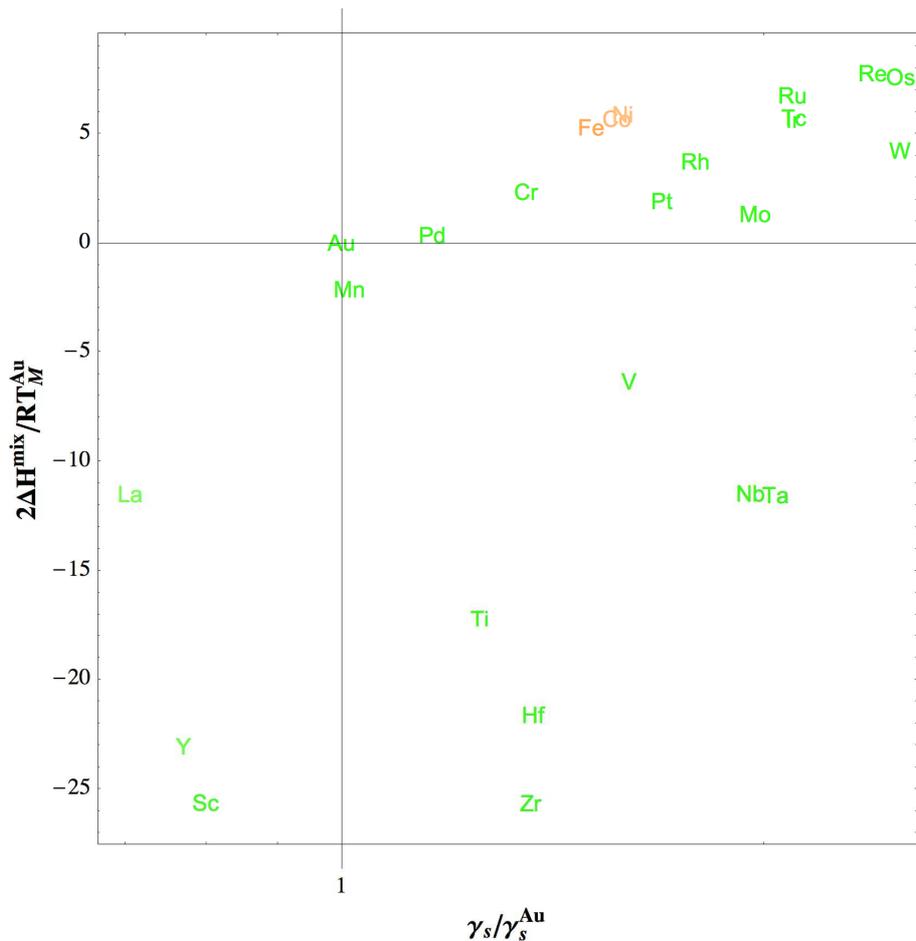

**Figure S5:** The same map as in Figures 7 and 8, but only for Au-based alloys to facilitate comparison with experiment. The map is colored in the same manner as Figure 8.

**Supplementary References:**